# Network-Scale Road Disruption from Liquefaction in Cascadia Subduction Zone Earthquakes

M.D. Sanger[1], O. Blaze-Smith[2], B.W. Maurer[3], L. Wotherspoon[4], M.O. Eberhard[5], and J.W. Berman[6]

## ABSTRACT

This paper presents a mechanics-informed, data-driven framework for modeling liquefaction-induced disruption of roadway networks following a magnitude-9 earthquake on the Cascadia Subduction Zone (CSZ). Liquefaction hazard is predicted using a geospatial liquefaction model trained on more than 37,000 cone penetration tests (CPTs) and conditioned on spatial data describing geomorphology, hydrology, climate, and surficial geology. Ground motions are derived from physics-based ensembles of CSZ rupture scenarios. Segment-level probabilities of closure and service degradation are estimated using empirically derived fragility relationships and propagated through the National Highway System using spatially correlated Monte Carlo simulation. Results indicate strong concentration of impacts in coastal lowlands, estuaries, and river valleys, with pronounced exposure along U.S. Route 101. Focused analysis in Pacific and Grays Harbor Counties, Washington, shows elevated probability of healthcare isolation. Compared with prior statewide assessments based on geologic screening, predicted bridge closures due solely to liquefaction are an order of magnitude lower, reflecting improved representation of subsurface conditions and damage mechanisms. Despite uncertainties, the framework provides a defensible tool for transportation resilience planning and asset prioritization in Cascadia.

## Introduction

The Cascadia Subduction Zone (CSZ) extends approximately 1,300 km from northern California to Vancouver Island and is capable of producing earthquakes of magnitude 8 to 9. Paleoseismic evidence indicates that the most recent full-margin rupture occurred in 1700 CE and that similar events have occurred repeatedly over the Holocene. The next CSZ earthquake is expected to cause severe shaking over large areas of the U.S. Pacific Northwest, with widespread secondary hazards including liquefaction, landslides, and tsunami inundation. Among these, liquefaction represents one of the most geographically extensive threats to infrastructure, particularly transportation systems that support emergency response and post-event recovery [1-3]. Previous regional assessments of liquefaction impacts on transportation networks have relied largely on geologic susceptibility classes and simple acceleration thresholds, mapping liquefaction as a binary outcome across broad map units. These methods provide a coarse screening tool but do not account for spatial variability in subsurface conditions or the mechanics governing soil behavior. As a consequence, these approaches may significantly overestimate impacts in some locations and underestimate them in others. This work introduces a regional framework that integrates geotechnical insight with data-driven prediction and spatial simulation to improve estimation of road-network disruption following a CSZ rupture.

[1] Dept. of Civil Engineering, University of Washington, Seattle, WA 98195 (sangermd@uw.edu)
[2] Dept. of Civil Engineering, University of Washington, Seattle, WA 98195 (smitho2@uw.edu)
[3] Dept. of Civil Engineering, University of Washington, Seattle, WA 98195 (bwmaurer@uw.edu)
[4] Dept. of Civil and Environmental Eng, University of Auckland, Auckland, NZ (l.wotherspoon@auckland.ac.nz)
[5] Dept. of Civil and Environmental Engineering, University of Washington, Seattle WA, USA (eberhard@uw.edu)
[6] Dept. of Civil and Environmental Engineering, University of Washington, Seattle WA, USA (jwberman@uw.edu)

## Data and Methodology

### Ground Motion Input

Ground motions are taken from the ensemble scenario simulations of Wirth et al. [4], who generated physics-based ShakeMaps for magnitude-9 ruptures on the CSZ using variability in rupture geometry, depth, and slip distribution. Median peak ground accelerations (PGA) are adopted to represent expected shaking intensity. This ensemble-based approach provides a more realistic representation of spatial shaking patterns than reliance on a single deterministic rupture or empirical attenuation models alone.

### Geospatial Liquefaction Model (GLM)

Liquefaction hazard is quantified using the Geospatial Liquefaction Model (GLM) of Sanger et al. [5], which was trained using a mechanics-informed machine learning (ML) approach, and which sought to avoid many of the common pitfalls with existing data-driven models [6]. The model uses supervised ML to infer subsurface conditions from geospatial predictors including slope, elevation, distance to water bodies, soil classification, surficial geology, and climatic proxies. Predictions are anchored to CPT-based liquefaction mechanics through use of the Liquefaction Potential Index (LPI), which estimates severity of surface manifestation as a function of cyclic demand and soil resistance. Regression kriging is applied to update predictions near available CPT measurements, ensuring physical consistency with field data.

### Fragility and Road Performance

Expected liquefaction manifestation severity is translated into road performance using fragility relationships developed by Geyin and Maurer [7], which provide probabilities of no, minor, moderate, or severe surface manifestation conditional on LPI. These damage states are mapped to probabilities of service reduction and closure based on empirical evidence from liquefaction case histories and analog hazards such as flooding and snow cover. Minor manifestation is assumed to rarely cause closure, whereas severe manifestation is assumed to render a road impassable with higher probability.

### Network Modeling

Road networks are discretized into approximately 90-m segments and grouped into links between intersections. Segment-level probabilities are propagated to links using correlated Monte Carlo simulation. A Gaussian copula is employed to preserve observed spatial correlation in liquefaction hazard, with correlation length estimated empirically from semivariograms of predicted damage probabilities. This approach prevents unrealistic independence assumptions that would overpredict network disruption.

## Results

Predicted LPI values exhibit strong geographic organization, with highest hazard concentrated in coastal lowlands, estuaries, and alluvial river valleys. Urban waterfronts such as Seattle, Olympia, and Portland show elevated hazard consistent with artificial fill and shallow groundwater. Segment-level results show that closures are rare in upland regions but highly clustered in specific coastal corridors. U.S. Route 101 is identified as the most critical vulnerable corridor in all three states. At the link scale, cumulative effects of small segment probabilities result in substantial connectivity loss along lengthy routes. County-scale modeling in Pacific and Grays Harbor Counties reveals high likelihood of isolation and loss of hospital access in several communities, particularly those served by single access routes. Predicted liquefaction impacts to bridges are substantially lower than estimates from earlier statewide screening studies. When closure probabilities are integrated across the bridge inventory, expected closures are approximately one order of magnitude fewer than previously reported, highlighting conservatism inherent in geologic screening methods and overly simplistic binomial thresholds for expected damage. Complete results are provided online as a package of GIS files and are described in detail in an accompanying manuscript [8].

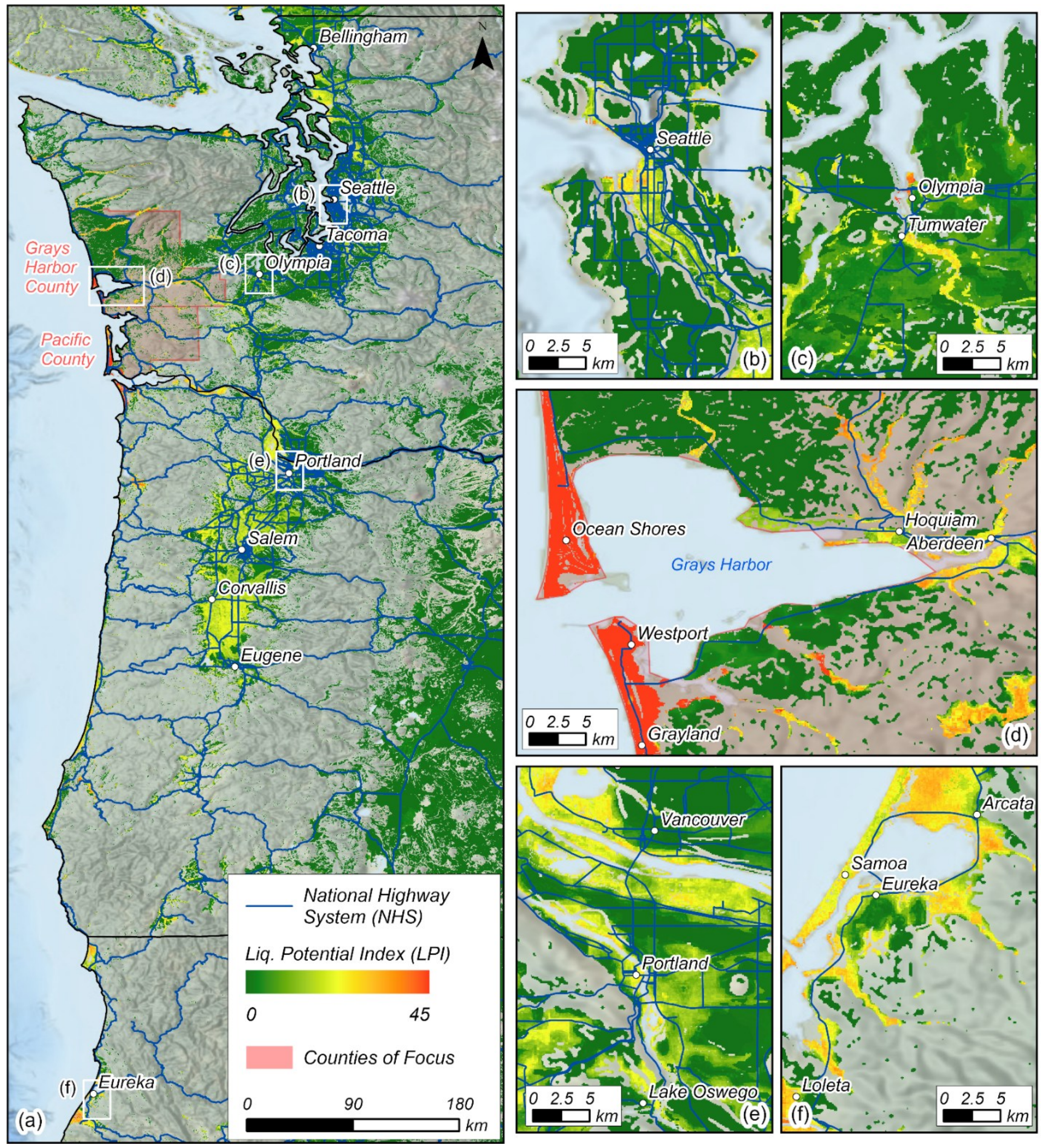

Figure 1. Predicted liquefaction potential index (LPI) in a median M9 CSZ rupture: (a) full study area; (b) Seattle, WA; (c) Olympia, WA; (d) Grays Harbor, WA; (e) Portland, OR; and (f) Eureka, CA.

**Socioeconomic Context**

Correlation analysis indicates that road disruptions are weakly but significantly associated with indicators of wealth, educational attainment, and age distribution. Lower-income communities and populations with higher age dependency ratios show higher expected exposure. These results suggest that liquefaction impacts may exacerbate social vulnerability patterns, reinforcing the need for equitable resilience planning.

## Conclusions

A regional framework for modeling liquefaction-induced transportation disruption has been presented. By integrating geotechnical mechanics with geospatial machine learning and network simulation, the approach substantially improves upon prior screening methodologies. Results revise expectations of bridge failure while highlighting acute vulnerability in coastal regions. The framework is scalable and adaptable for application to other regions and hazards and provides a practical tool for prioritization and resilience investment in Cascadia.

## Acknowledgments

The presented work is based on research supported by the United States Geological Survey (USGS) under award G23AP00017, the Cascadia Region Earthquake Science Center (CRESCENT) via National Science Foundation (NSF) award 2225286, the Cascadia Coastlines and Peoples (CoPes) Hub via NSF award 2103713, the Pacific Earthquake Engineering Research (PEER) Center under award 1185-NCTRMB, and the Pacific Northwest Transportation Consortium (PacTrans) under award 69A3552348310. However, any opinions, findings, conclusions, or recommendations expressed herein are those of the authors and may not reflect the views of USGS, NSF, CoPes, CRESCENT, PEER, or PacTrans.